\documentclass[aps,prd,nofootinbib,superscriptaddress,onecolumn,11pt,notitlepage]{revtex4-2}

\usepackage{xcolor}
\usepackage{url}
\usepackage[hidelinks]{hyperref}
\usepackage{amssymb,amsmath}
\usepackage[locale=US, exponent-product=\cdot]{siunitx}
\usepackage{graphicx}

\begin{document}
\raggedbottom

\title{Opportunities for Gravitational Wave Physics at the South Pole}

\author{C.~A.~Arg\"uelles}
\affiliation{Department of Physics \& Laboratory of Particle Physics and Cosmology, Harvard University, 18 Oxford St., Cambridge, MA 02138, USA}

\author{M.~DuVernois}
\affiliation{Wisconsin IceCube Particle Astrophysics Center \& Department of Physics, University of Wisconsin--Madison, 222 West Washington Ave., Suite 500, Madison, WI 53703, USA}

\author{P.~W.~Graham}
\affiliation{Leinweber Institute for Theoretical Physics \& Kavli Institute for Particle Astrophysics and Cosmology, Department of Physics, Stanford University, Stanford, CA 94305, USA}

\author{T.~Kovachy}
\affiliation{Department of Physics and Astronomy, Center for Fundamental Physics, \& Center for Interdisciplinary Research and Exploration and Research in Astrophysics, Northwestern University, 2145 Sheridan Road, Evanston, IL 60208, USA}

\author{J.~Mitchell}
\affiliation{Cavendish Laboratory, University of Cambridge, J.J. Thomson Ave., Cambridge CB3 0HE, United Kingdom}

\begin{abstract}
Atom interferometers represent a promising approach for gravitational wave detection in the decihertz frequency band, complementary to existing light-based detectors.
The South Pole offers unique advantages for such experiments: exceptionally low seismic noise, established infrastructure for large scientific projects, and a location that strengthens gravitational wave source localization through global triangulation.
Here we discuss the scientific case and practical considerations for deploying a long-baseline atom interferometer at the South Pole, which has the potential to expand the global network of gravitational wave detectors while enabling precision tests of fundamental physics.
\end{abstract}

\maketitle

\begin{figure}[h!]
\centering
\includegraphics[width=1.0\textwidth]{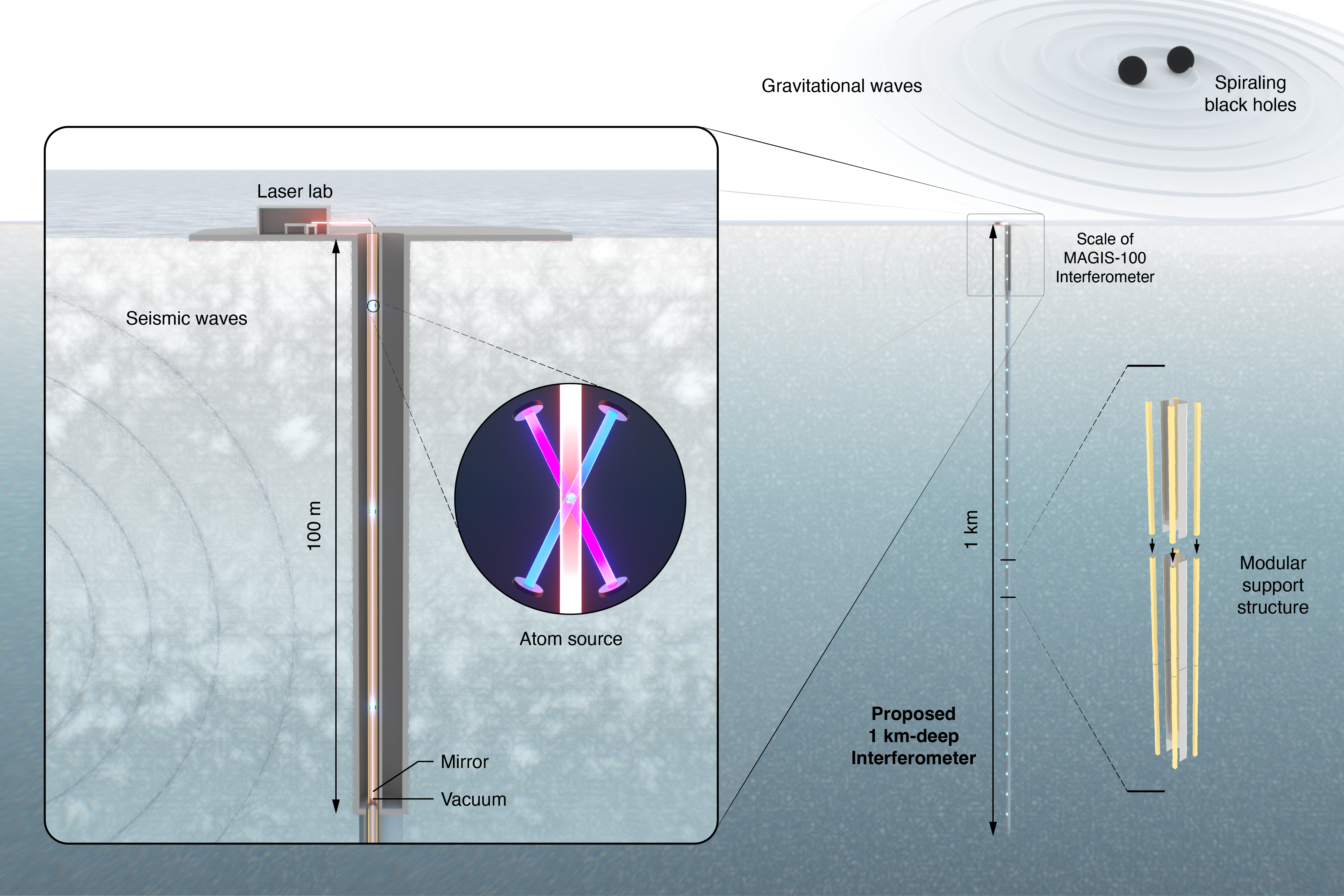}
\caption{Conceptual illustration of a kilometer-scale vertical atom interferometer at the South Pole. A vacuum tube extending approximately 1\;km into the Antarctic ice sheet houses the interferometer baseline, with a laser laboratory at the surface, atom sources along the vertical baseline, and a retroreflecting mirror at the bottom. This represents a tenfold scale-up from the MAGIS-100 detector currently under construction at Fermilab. The surrounding ice provides natural thermal stability and vibration isolation for the instrument.}
\label{fig:placeholder}
\end{figure}

Long-baseline atom interferometers offer a promising approach to terrestrial gravitational wave detection in the mid-band frequency range, roughly 0.3\;Hz to 3\;Hz~\cite{dimopoulos2008atomic,graham2013new,yu2011gravitational} and are currently supported by an international community effort~\cite{abdallaTerrestrial:2025}. This band lies below the reach of current generation ground-based laser interferometers such as LIGO~\cite{Abbott2016}, which are limited by seismic noise at frequencies below a few hertz, and above the millihertz band at which the planned space-based LISA observatory will have its peak sensitivity~\cite{prince2002lisa}. The mid-band is scientifically rich: black hole and neutron star binaries radiate in this range for hours to days before merger, providing early-warning alerts that would allow electromagnetic and neutrino telescopes to observe the final coalescence in real time~\cite{graham2018localizing}. This multi-messenger capability would be transformative---for example, advanced localization of neutron star mergers could enable observation of the prompt emission that current detectors miss entirely. The mid-band range may also contain gravitational wave backgrounds of cosmological origin, including signals from first-order phase transitions in the early universe at energy scales above the reach of particle colliders~\cite{Caprini:2019egz}, and from other exotic sources such as cosmic strings and axion dynamics~\cite{graham2017mid}.

Realizing such a detector requires a site that combines low seismic noise, favorable location, and the infrastructure to support large-scale scientific construction. The Amundsen-Scott South Pole Station---home to the IceCube Neutrino Observatory and the South Pole Telescope---offers all three. The South Pole's seismic environment is among the quietest on Earth~\cite{anthony2021sixdecades,richter2026quietest}, the vertical orientation of an interferometer there naturally suppresses systematic uncertainties that arise from Coriolis forces, and a specially managed quiet zone surrounds the station. Beyond gravitational wave science, such an instrument would enable precision tests of the equivalence principle, searches for new fundamental forces, and probes of wavelike dark matter~\cite{abend2024terrestrial}. Deployed as part of a global network of mid-band detectors, a South Pole interferometer would provide critical southern-hemisphere coverage for source localization on the sky.

\section*{Why the South Pole?}

The Amundsen-Scott South Pole Station, one of three permanent US Antarctic stations operated by the National Science Foundation, supports international science year-round with a winter crew of roughly 40 and a summer population of 120--250.
The station is resupplied by US military ski-equipped LC-130 aircraft and three annual overland traverses---tractors hauling large sleds of fuel and cargo from the coastal logistics base at McMurdo Station.
The ability to deliver large volumes of construction material has been demonstrated repeatedly: the IceCube Neutrino Observatory alone required approximately 9.5 million pounds of cargo and fuel, delivered over about 300 LC-130 flights.
A new contract for ski-equipped aircraft is now addressing the aging LC-130 fleet.

Installing a kilometer-scale vertical interferometer would draw on drilling expertise developed for IceCube, which used a hot-water drill to bore 86 holes to depths of 2.5\;km in the ice sheet, with 6 additional holes drilled in the 2025-2026 field season for the IceCube Upgrade.

The station's existing power plant, communications infrastructure, Austral Summer logistics, and year-round support personnel provide a foundation for sustained operation.
Establishing a detector within the designated quiet zone surrounding the station would further reduce anthropogenic noise sources.
Although the glacier moves about 9m per year, shear is only observed well below the top kilometer of ice~\cite{doi:10.1073/pnas.082238999} and has not adversely affected the IceCube instrumentation down to nearly 2500m below the surface since their installation in 2004--2011.

The precedent for large scientific projects at the South Pole is well established.
Before the South Pole Telescope and IceCube, research at the South Pole was modest in scale, spanning seismometry, cosmic rays, glaciology, and atmospheric monitoring.
Constructing those flagship experiments required substantial expansions of logistics capability, driven by compelling science cases for cosmic microwave background measurements and neutrino astronomy.
A gravitational wave detector, such as the one proposed here, would be synergistic with the further development of South Pole logistics, e.g., future generator power upgrades, increases in communications bandwidth, and installation of next-generation astronomical instruments.

\section*{Design Concepts}

The detector concept, illustrated in Fig.~\ref{fig:placeholder}, centers on a vertical atom interferometer with an approximately 1\;km baseline housed within a borehole drilled into the Antarctic ice sheet. The basic architecture follows the gradiometer approach developed for MAGIS-100~\cite{abe2021matter}, scaled up by an order of magnitude.
A laser laboratory at the surface generates the optical pulses that drive single-photon transitions on a narrow atomic line (such as the clock transition in strontium), where multiple atom sources at different depths in the ice launch ultracold atom clouds into free fall along the vertical baseline.
The different clouds interact with the laser at different depths.
A retroreflecting mirror at the bottom of the shaft returns the laser beam, so that alternating single-photon transitions from pulses propagating in opposite directions can be used to boost the instrument's sensitivity.  Differential phase measurements between spatially separated atom clouds are sensitive to gravitational wave strain, which modifies the light travel time between atom clouds, while suppressing residual laser noise \cite{graham2013new}.

The vacuum tube running the length of the borehole would require a modular support structure compatible with installation in stages and flexible interfacing with experimental subsystems, similar to the segmented deployment of IceCube's digital optical modules on kilometer-length cables.
The surrounding ice sheet provides a thermally stable environment~\cite{doi:10.1073/pnas.082238999} and natural isolation from surface vibrations, both of which benefit interferometer performance.

At 1\;km, this baseline would be ten times longer than MAGIS-100 and comparable to proposals for km-scale detectors at other sites~\cite{abend2024terrestrial}.
The longer baseline directly improves gravitational wave strain sensitivity, increasing the detector's reach in the astrophysically rich decihertz band where compact binary inspiral signals are strongest.
Detailed engineering studies would be required to establish borehole diameter, vacuum requirements, and thermal management.


\section*{Experimental Advantages and Expected Sensitivity}

We now examine in detail how the South Pole's site characteristics translate into concrete improvements in detector sensitivity and scientific reach.

\begin{figure}[h!]
\centering
\includegraphics[width=0.8\textwidth]{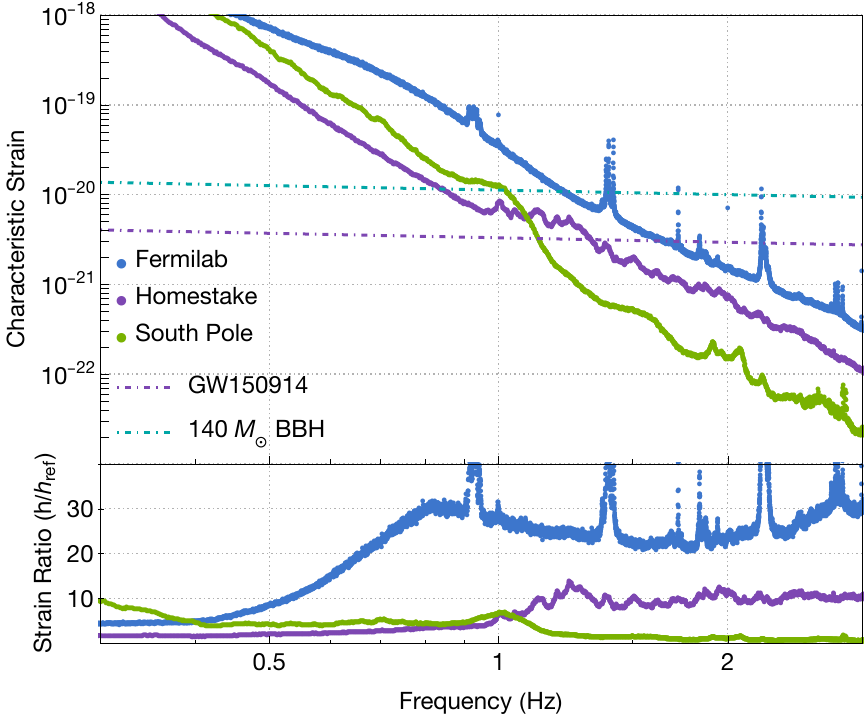}
\caption{Gravity gradient noise (GGN) limits on characteristic strain sensitivity. Comparison for a model vertical 1\;km baseline atom interferometer, derived from broadband seismic data. Gravity gradient noise models are based on Rayleigh waves~\cite{ggn_model_notes,harmsTerrestrial:2019,mitchellMAGIS100:2022,badurinaUltralight:2023}. Seismic data was acquired through the IRIS database and analyzed through standard methods~\cite{anthony2021sixdecades,mandic3D:2018a}. Top: characteristic strain noise at Fermi National Accelerator Laboratory (site of MAGIS-100), the South Pole (IRIS IU QSPA), and Homestake, South Dakota (IRIS IU RSSD), with two example binary black hole merger signals shown as dashed lines, GW150914 is the first detection by LIGO~\cite{Abbott2016} and $140~M_\odot$ is a theoretical 140-140 solar mass binary black hole merger. Bottom: ratio of modeled GGN strain between site seismic data and reference data from the Peterson New Low Noise Model~\cite{petersonObservations:1993} (lowest possible ground motion globally), illustrating the South Pole's exceptionally quiet seismic environment.}
\label{fig:seismic-compare}
\end{figure}

\textbf{Seismic noise and gravity gradient noise.}
In atom interferometry with freely falling atoms, the atoms have no mechanical connection to the Earth, and free fall in ultrahigh vacuum significantly isolates them from seismic vibrations.
Laser vibrations induced by seismic waves can be suppressed through differential measurements between atom interferometers separated over a long baseline and controlled by the same laser beam~\cite{yu2011gravitational,graham2013new}.
However, seismic waves still couple to the atoms via Newtonian gravity gradient noise (GGN): density fluctuations in the surrounding medium produce time-varying gravity gradients that constitute a prominent noise background~\cite{harmsTerrestrial:2019}.
Mitigation strategies for GGN have been proposed~\cite{miga:2018,mitchellMAGIS100:2022,badurinaUltralight:2023} but remain experimentally undemonstrated.
Choosing a site with intrinsically low seismic noise is therefore critical.
As shown in Fig.~\ref{fig:seismic-compare}, the South Pole is exceptionally quiet: between 0.5 and 1\;Hz, the GGN-limited strain sensitivity at the South Pole is 3--5 times better than at the Fermilab site, and above 1\;Hz it is more than 30 times better than Fermilab and nearly 10 times better than the Homestake mine---one of the deepest mines in the United States.

\begin{figure}[h!]
\centering
\includegraphics[width=\textwidth]{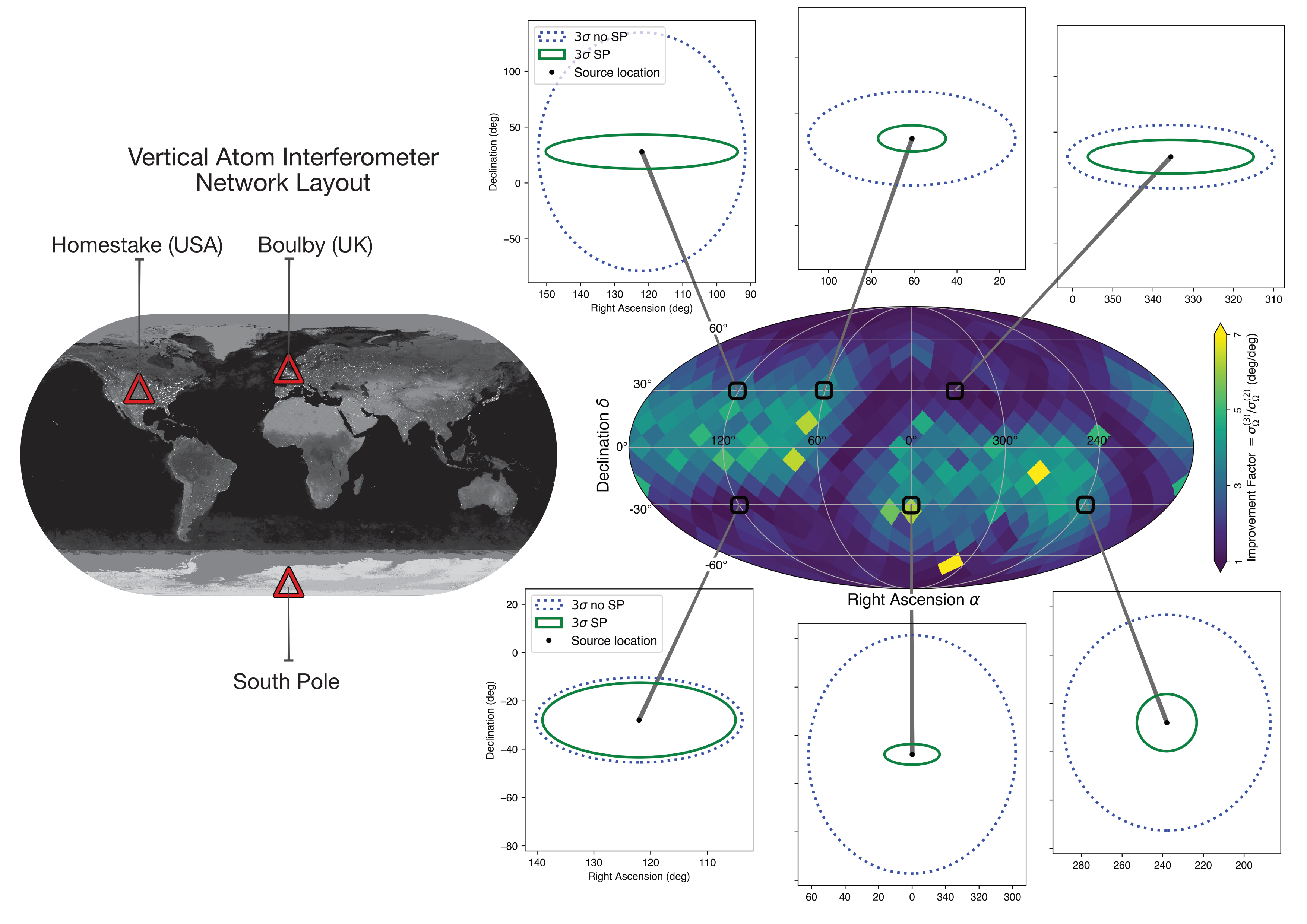}
\caption{Sky-localization improvement from adding a South Pole detector to a two-site northern-hemisphere network (Homestake (SURF), Boulby)~\cite{skymap_notes}. Left: Graphical layout of the three detector sites. Right: sky map of the fractional improvement in angular resolution (colorbar: ratio of angular resolution improvement between 3 detector and 2 detector networks). Simulation used open source project \href{https://github.com/sbaum90/AIMforGW}{AIMforGW}~\cite{baumGravitational:2024}. Top and bottom: six example sources showing 99.7\% confidence regions without (blue ellipse) and with (green ellipse) the South Pole detector; dots mark the true source direction.}
\label{fig:sky-localization}
\end{figure}

\textbf{Sky localization.}
One significant potential benefit of observing gravitational waves in the mid-band between LIGO and LISA is excellent angular localization of most sources~\cite{graham2018localizing}.
Using a geographically distributed network of detectors improves the angular resolution.
Current and proposed projects for long-baseline atom interferometers over 100~m are concentrated in the northern hemisphere---at Fermilab (MAGIS-100)~\cite{abe2021matter} in the US, in the UK (AION)~\cite{Badurina_2020}, in Europe (MIGA, ELGAR, AICE)~\cite{miga:2018,ELGAR:2020,baynham2025letter}, and in China (ZAIGA)~\cite{ZAIGA:2020}.
The South Pole instrument is both geographically distant from the other detectors and also aligned along a unique direction (the Earth's rotation axis), thus improving the angular resolution of the network.
Fig.~\ref{fig:sky-localization} shows the improvement in sky localization obtained by adding a South Pole detector to a two-site northern-hemisphere network (Homestake (SURF), Boulby)~\cite{abdallaTerrestrial:2025}, computed using a Fisher-matrix analysis of the detector response functions~\cite{graham2018localizing,baumGravitational:2024}.
The inclusion of the South Pole instrument improves source localization across large fractions of the sky, for some source locations up to an order of magnitude.
And the angular resolution becomes more uniform across the sky as the South Pole detector removes some areas where a two-detector network did not have good coverage.

\textbf{Coriolis forces.}
Coriolis forces arising from the coupling of atomic motion to Earth's rotation are a major challenge for atom interferometers. When the rotation vector and interferometer axis are not parallel, Coriolis forces prevent the two branches of the atomic wavefunction from perfectly overlapping, reducing fringe contrast and introducing velocity-dependent phase noise~\cite{lanCoriolis:2012a,dickersonMultiaxis:2013}.
Rotation compensation techniques exist~\cite{lanCoriolis:2012a,dickersonMultiaxis:2013}, but the large lateral beam displacement over a km-scale baseline makes their practicality uncertain~\cite{glick2024coriolis,abend2024terrestrial}. At the South Pole, a vertical interferometer axis is naturally parallel to Earth's rotation vector, suppressing these Coriolis effects without active compensation.


For gravitational wave detection specifically, multiloop interferometer geometries~\cite{dubetsky2006atom,wang2024robust} that mitigate Coriolis effects will likely be needed anyway to suppress noise from atom cloud jitter coupling to static gravity gradients~\cite{abend2024terrestrial,abe2021matter}. The definitive Coriolis advantage of the South Pole therefore, lies primarily in other science targets---tests of the equivalence principle, searches for new fundamental forces, and probes of low-frequency wavelike dark matter~\cite{abend2024terrestrial}---which involve static or slowly varying signals incompatible with multiloop configurations. The South Pole thus significantly broadens the scientific reach of a next-generation long-baseline atom interferometer.


\section*{Conclusion}

The South Pole has repeatedly proven to be a uniquely enabling site for frontier science: its transparent, low-radioactivity ice made possible the IceCube Neutrino Observatory, its dry and stable atmosphere enabled transformative cosmic microwave background measurements with the South Pole Telescope, and here we highlight yet another way in which this location is exceptional.
The combination of intrinsically low seismic noise, a vertical geometry that naturally suppresses systematic uncertainties from Coriolis forces, and decades of proven logistics infrastructure makes the South Pole a potentially compelling site for a next-generation long-baseline atom interferometer that merits further study.

Several important questions remain to be addressed, including a characterization of atmospheric gravity gradient noise at the South Pole and detailed site surveys, borehole engineering and ice studies, a conceptual design for the vacuum and cold-atom systems, and studies of strategies for installation and maintenance.
These investigations would build on the extensive body of glaciological and geophysical data already collected at the Pole and could be synergistic with the development of other next-generation projects such as IceCube-Gen2.

Looking ahead, a South Pole atom interferometer deployed as part of a global network of mid-band gravitational wave detectors may offer opportunities to open the decihertz band to multi-messenger astronomy while enabling precision tests of fundamental physics at unprecedented sensitivity.

\begin{acknowledgments}
We thank Jan Rudolph for valuable discussions. CAA are supported by the Faculty of Arts and Sciences of Harvard University, the National Science Foundation, the John Templeton Foundation, the Research Corporation for Science Advancement, the Canadian Institute for Advanced Research, and the David \& Lucile Packard Foundation.
TK acknowledges support from the Gordon and Betty Moore Foundation and the David \& Lucile Packard Foundation.
JM is supported by the UKRI STFC grant numbers ST/T006579/1, ST/W006200/1, ST/X004864/1, and the Isaac Newton Trust.
MD is supported by the National Science Foundation, the University of Wisconsin--Madison, and the Wisconsin Alumni Research Foundation.
We thank Leidos as the Antarctic Support Contractor for the United States Antarctic Program.
PG was supported in part by NSF Grant No. PHY-2310429, Simons Investigator Award No. 824870, and the John Templeton Foundation Award No. 63595.
Claude, an AI assistant developed by Anthropic, was used to edit and review the text under a physicist's supervision.
\end{acknowledgments}

\noindent\textbf{Competing Interests.} The authors declare that they have no competing financial interests.

{\sloppy
\noindent\textbf{Correspondence.} Correspondence and requests for materials should be addressed to C.~A.~Arg\"uelles~(email: \href{mailto:carguelles@g.harvard.edu}{carguelles@g.harvard.edu}).
\par}

\bibliography{magis-at-sp}

\begin{thebibliography}{32}
\bibitem [{\citenamefont {{Methodological
  note}}({\natexlab{b}})}]{skymap_notes}%
  \BibitemOpen
  \bibfield  {author} {\bibinfo {author} {\bibnamefont {{Methodological
  note}}},\ }\href@noop {} {\bibinfo {title} {Sky localization fisher-matrix
  analysis}} ({\natexlab{b}}),\ \bibinfo {note} {sky localization used Fisher
  information analysis method for a network of 1\;km vertical baseline atom
  interferometers. Model binary black hole source parameters used chirp mass of
  25\;$M_{\odot}$, luminosity distance 100\;Mpc, binary inclination angle of
  $\pi/4$ and polarization angle $\pi/3$. The Fisher information was integrated
  over the mid-band frequency range. Simulation used a detector noise model
  based on an optimal Homestake GGN floor with a factor of 10 assumed noise
  suppression.}\BibitemShut {Stop}%
\end{thebibliography}

\end{document}